\documentstyle{article}

\begin{document}
\title{Coexistence of Quantum Theory and Special Relativity in signaling scenarios}
\author{Pedro Sancho \\ GPV de Valladolid \\ Centro Zonal
en Castilla y Le\'on \\ Ori\'on 1, 47014, Valladolid, Spain }
\date{}
\maketitle
\begin{abstract}
The coexistence between Quantum Mechanics and Special Relativity is
usually formulated in terms of the no-signaling condition. Several
authors have even suggested that this condition should be included
between the basic postulates of Quantum Theory. However, there are
several scenarios where signaling is, in principle, possible: based
on previous results and the analysis of the relation between
unitarity and signaling we present an example of a two-particle
interferometric arrangement for which the dynamics is, in principle,
compatible with superluminal transmission of information. This type
of non-locality is not in the line of Bell's theorem, but closer in
spirit to the one-particle acausality studied by Hegerfeldt and
others. We analyze in this paper the meaning of this non-locality
and how to preserve the coexistence of the two fundamental theories
in this signaling scenario.
\end{abstract}
\section{Introduction}
According to most theorists the conflict between Quantum
Mechanics and Special Relativity due to the existence of nonlocal
correlations does not represent a fundamental difficulty for the
coexistence of both theories because of the no-signaling
condition \cite{Shi,Ghi,Ebe}. This condition prevents the
transmission of superluminal signals.

Recently it has been discussed by several authors the possibility
of including the no-signaling condition or some related axiom
between the basic postulates of Quantum Theory \cite{Sim,Sal}. In
particular, it has been explored the possibility of deriving the
linearity of the theory from a set of assumptions which includes
the prohibition of superluminal communication \cite{Sim}. An
example of a system which, at least in principle, could signal
would drastically restrict  the scope of these programmes and,
more important, would undoubtedly indicate the existence of
physical mechanisms beyond the quantum dynamics of the system
responsible for the coexistence of Quantum Mechanics and Special
Relavity.

We present in this paper an example of this type. As we shall see
the possibility of dynamic signaling (the signaling when only the
dynamic evolution ruled by Schr\"odinger's equation is taken into
account without including considerations about emission, detection
or other types of processes) easily follows from the existence of
one- particle interferences in completely entangled two-particle
systems. In previous work it has been found that for some types of
interferometric arrangements of completely entangled particles in
momentum we can have one-particle interferences \cite{SPRa, Spa,
Spl}. These interferences are not excluded by the results of Ref.
\cite{Jae}, where it was demonstrated the absence of one-particle
interferences for a large class of completely entangled systems.

Of course, we do not want to say that these effects are
experimentally accessible. Probably their magnitude is exceedingly
small to be experimentally measured. A simple estimation based on
the results of Refs. \cite{Spa,Spl} shows the extremely thin
magnitude of the interferences. Moreover, as we shall discuss
later, other conditions of a non-dynamic type should be fulfilled
in order to observe these effects. However, although they surely
are inaccessible to the experiment their conceptual impact can be
important. For instance, they say to us that the dynamics is
compatible with signaling. These considerations are very close in
spirit to the analysis by Hegerfeldt and others of the acausal spread of
one-particle wavefunctions with initial localization \cite{Heg}.
It was shown that systems which are approximately localized at a
given time, spread at later times in a way that violates
Einstein's causality. In words of the author,
\begin{quotation}
this possible acausality is seen more as a problem of the
underlying theory than as an experimentally verifiable prediction
\end{quotation}
In a related context other authors have discussed the possible
effects associated with the non-vanishing of the Feynman
propagator in space-like regions \cite{Rub, Svo}. In particular,
it has been suggested that this property could be used to generate
entanglement in a superluminal way (although it cannot be used to
transmit information faster than the speed of light) \cite{Fra}.
Finally, we must refer to the paper \cite{Aha}, where some aspect
of non-locality no conflicting with the non-signaling condition
are discussed.

In order to go deeply into the physical meaning of this new type of
non-locality, which is not within the usual framework provided by
Bell's theorem, we shall emphasize the relation between unitarity
and locality. We shall show that the non-locality associated with
interferometric arrangements is closely linked to an effective (not
from a fundamental type)non-unitarity  that emerges when we restrict
our considerations to only the detected particles disregarding those
absorbed by the screen. We shall demonstrate this property in the
path integration formalism.

Assuming the impossibility of superluminal communication as a
fundamental constrain on any physical system, we must explore the
physical mechanisms that prevent in our arrangement the well- known
causal paradoxes associated with the violation of that postulate of
Special Relativity. We shall show that in addition to the dynamic
condition for the transmission of superluminal signals another
conditions (not from a dynamic type) must be fulfilled. In this
paper, as an example, we shall consider a condition associated with
the rate of emission of entangled particles by the source. Other
conditions of non- dynamic type have been considered in the
literature. In particular in Ref. \cite{Ste} it has been shown that
the detection process preserves the cause-effect order for fast
light (pulses with superluminal group velocity).

The plan of the paper is as follows. In Sect. 2 we present the
dynamic no-signaling condition in terms of the reduced density
matrix of the system. We show that this condition is not
fulfilled
by some interferometric arrangements as those discussed in Refs.
\cite{SPRa, Spa, Spl}. The relation between unitarity and
locality is discussed in Sect. 3, where we also demonstrate that
the evolution in some interferometric arrangements is non-unitary
in an effective way. Section 4 deals with a constrain on
the ratio of emission of entangled pairs by the source necessary
to preserve the no-signaling. Finally, in the discussion we
consider the results obtained in the paper.

\section{Signaling and one-particle interferences in two-particle systems}

We show in this section that the existence of one-particle
interferences in completely entangled two-particle systems would
provide (in absence of other inhibiting mechanisms) a simple way,
at least in principle, for signaling.

We consider the following ideal experiment. A source emits pairs of
particles that fall upon two diffraction gratings with two slits
each. Slits A and B are in one side and C and D in the other. Slit C
is just placed in front of A and D of B. The particles are prepared
in the maximally entangled state (we only consider the stationary
problem):
\begin{equation}
\psi ({\bf x_1}, {\bf x_2}) = \frac{1}{\sqrt{2}} \psi _{1A} ({\bf
x_1}) \psi _{2D} ({\bf x_2}) + \frac{1}{\sqrt{2}} \psi _{1B} ({\bf
x_1}) \psi _{2C} ({\bf x_2}) \label{eq:sta}
\end{equation}
For instance, we can have a source that emits pairs of particles
completely correlated in momentum,
\begin{equation}
{\bf p_1}+{\bf p_2}=0
\label{eq:cruz}
\end{equation}
with ${\bf p_1}$ and ${\bf p_2}$ the momenta of the two particles.
Obviously, in quantum theory the pairs cannot be prepared in this
state because of the uncertainty relations, that would imply a
complete delocalization of the (center of mass of the) source.
However, reaching an adequate compromise between the localization
of the (center of mass of the) source and the deviations with
respect to the exact relation (\ref{eq:cruz}) we can have states
of the type (\ref{eq:sta}) to a good approximation (see, for
instance, Ref. \cite{Spl} for a source of finite size). We shall
return to this point later in the Section.

The discussions about one-particle interferences in entangled
systems are usually carried out in terms of the
reduced density matrix, which for particle 1 is defined by
\begin{eqnarray}
\rho _1 ({\bf x_1}, {\bf x'_1})=\int d^3 {\bf x_2} \psi ({\bf x_1},
{\bf x_2}) \psi ^* ({\bf x'_1}, {\bf x_2})= \frac{1}{2}
\psi _{1A} ({\bf x_1}) \psi _{1A}^* ({\bf x'_1}) + \nonumber \\
\frac{1}{2} \psi _{1B} ({\bf x_1}) \psi _{1B}^* ({\bf x'_1}) +
\frac{1}{2} I \psi _{1A} ({\bf x_1}) \psi _{1B}^* ({\bf x'_1})
+ \frac{1}{2} I^* \psi _{1B} ({\bf x_1}) \psi _{1A}^* ({\bf
x'_1})
\end{eqnarray}
where
\begin{equation}
I= \int d^3 {\bf x_2} \psi _{2D} ({\bf x_2}) \psi _{2C}^* ({\bf
x_2})
\end{equation}
is the scalar product of the two wave functions $\psi _{2C}$ and
$\psi _{2D}$ (we assume both to be normalized to unity).

Taking ${\bf x'_1}={\bf x_1}$ in the reduced density matrix we
see that the condition for the existence of one-particle
interferences is the scalar product to be different from zero.

The existence of one-particle interferences in the two-slit
experiment has been demonstrated in a number of ways:
semiclassical approximation \cite{SPRa}, Fresnel's functions and
Gaussian-slit approximation \cite{Spa} and Gaussian-slit
approximation with extended incoherent sources \cite{Spl}. These
results are in marked contrast with the demonstration of the absence
of one-particle interferences for a large class of completely
entangled states \cite{Jae}. As discussed at length in
\cite{SPRa},
the difference between both behaviors lies in the impossibility
in the first case of distinguishing between the available
alternatives for the particle using measurements in the companion
particle. In particular, the extended Feynman rule (following
the nomenclature of Ref. \cite{SPRa}) valid for the types of
arrangements considered in \cite{Jae} does not hold for those
considered here.

We note by completeness that the results of \cite{SPRa,Spa} could
be criticized because the particles stopped at the diffraction
grating were not taken into account. However, in \cite{Spl} these
stopped particles were included in the calculations without
destroying the one-particle interferences. The results of
\cite{SPRa,Spa, Spl} could also be criticized because of the use
of exactly opposite paths for both particles (a consequence in the
path integration formalism of the use of relations of the type
${\bf p_1}+{\bf p_2}=0$). As discussed before that exact match
between the momenta is incompatible with the total momenta-source
position uncertainty relations. However, the existence of paths
that are not exactly opposite gives rise to terms in Eq.
(\ref{eq:sta}) of the type $\psi _{1A} \psi _{2D}$ and $\psi _{1B}
\psi _{2C}$ or paths stopped by the screen (as those discussed in
Ref. \cite{Spl}). These new terms would reduce the visibility of
the fringes associated with the terms present in Eq.
(\ref{eq:sta}), but without eliminating their effects. Therefore,
we can continue our reasoning using only Eq. (\ref{eq:sta})
(although taking into account that this approach gives a
visibility larger than the real one).

Let us consider now the question of signaling. We introduce, for
instance in slit $C$, some physical element producing a phase
shift $\phi $ to the particles passing through that slit.
Therefore, after the slit the wave function is modified as
\begin{equation}
\psi _{2C} \rightarrow e^{i \phi }\psi _{2C}
\label{eq:ang}
\end{equation}
Introducing the polar decomposition of the wave functions, $\psi
_{\alpha } =R_{\alpha } e^{i \varphi _{\alpha }}$, we obtain for
the detection probability of particle 1
\begin{equation}
\rho _1  =\frac{1}{2} R_{1A}^2 + \frac{1}{2} R_{1B}^2
+ R_I R_{1A} R_{1B} cos (\varphi _I + \varphi _{1A} - \varphi
_{1B} - \phi )
\end{equation}
This expression clearly shows a displacement $ \phi $ of the one-
particle interference pattern in the opposite side.

This is an explicit example of nonlocality going beyond the
nonlocal correlations obtained in Bell-type experiments (see the
Discussion). We shall refer to it as a "dynamic signaling"
because it is only related to the dynamics or evolution of the
particles without taking into account other processes such as
emission or detection. We shall demonstrate in Sect. 4
that the dynamic signaling is only a necessary condition for
superluminal communication, which only becomes a sufficient one
when some additional conditions are imposed on the emission and detection
properties of the source.

\section{Unitarity and locality}

In this section we shall analyze in detail the relation between
unitarity and locality and to show that the the evolution in
interferometric experiments where we restrict our considerations
to the detected particles is non-unitary in an effective way.

\subsection{An example}

In order to simplify the analysis as much as possible we shall
carry out it in a simple state. Initially, at time $t_o$, the two
particles $1$ and $2$ are prepared in the state
\begin{equation}
|t_o > = \frac{1}{\sqrt{2}} |1_+ (t_o)> |2_- (t_o) > +
\frac{1}{\sqrt{2}} |1_- (t_o)> |2_+ (t_o) >
\end{equation}
where the $\pm$ subscripts refer to the possible values of a
binary observable, for instance the third component of a spin-1/2
particle.

The evolution of the state between times $t_o$ and $t$ is ruled
by the evolution operators $\hat{U} _1(t,t_o)$ and $\hat{U}
_2(t,t_o)$. We assume no interaction between both particles after
the initial preparation and, consequently, the total evolution
operator factorizes into the product of $\hat{U}_1$ and
$\hat{U}_2$. Then $|t_o >$ evolves into $|t >$, given by:
\begin{equation}
|t > = \frac{1}{\sqrt{2}} |1_+ (t)> |2_- (t) > +
\frac{1}{\sqrt{2}} |1_- (t)> |2_+ (t) >
\end{equation}
where $ |1_+ (t)> = \hat{U} _1(t,t_o) |1_+ (t_o)>,...$

The probability of detecting the particle $1$ in state $"+"$ at
time $t$ is
\begin{equation}
P(1_+(t)) = |<1_+(t)|<2_+(t)|t>|^2 + |<1_+(t)|<2_-(t)|t>|^2
\end{equation}
We rewrite this expression in terms of the evolution operators
and the initial states (which we assume to be orthonormal,
$<1_{\pm}(t_o)|1_{\pm}(t_o)>=1=<2_{\pm}(t_o)|2_{\pm}(t_o)>$ and
$<1_+(t_o)|1_-(t_o)>=0=<2_+(t_o)|2_-(t_o)>$). Assuming, moreover,
that $\hat{U}_1$ is unitary we obtain
\begin{eqnarray}
P(1_+(t)) = \frac{1}{2} |<2_+(t_o)| \hat{U}_2^+ (t,t_o) \hat{U}_2
(t,t_o) | 2_-(t_o)>|^2 + \nonumber \\
\frac{1}{2} |<2_-(t_o)| \hat{U}_2^+ (t,t_o) \hat{U}_2 (t,t_o) | 2_-(t_o)> |^2
\end{eqnarray}

Similarly, for $1_-$ we have
\begin{eqnarray}
P(1_-(t)) = \frac{1}{2} |<2_+(t_o)| \hat{U}_2^+ (t,t_o) \hat{U}_2
(t,t_o) | 2_+(t_o)>|^2 + \nonumber \\
\frac{1}{2} |<2_-(t_o)| \hat{U}_2^+ (t,t_o) \hat{U}_2 (t,t_o) | 2_+(t_o)> |^2
\end{eqnarray}
When $\hat{U}_2$ is also unitary these probabilities become
$P(1_+(t))=1/2$ and $P(1_-(t))=1/2$. The
probabilities on side $1$ are independent of the operations done
on side $2$. It is impossible to signaling using this
arrangement.

On the other hand, when $\hat{U}_2$ is not unitary, $\hat{U}_2
^+ \hat{U}_2 \neq \hat{1}$, the signaling becomes possible.
For instance, if we take
\[ \hat{U}_2^+ \hat{U}_2 = \left( \begin{array}{cc}
\alpha & \beta \\ \gamma & \delta \end{array} \right) \neq \hat{1} \]
the probabilities become
\begin{equation}
P(1_+(t))=\frac{1}{2} (|\beta|^2 + |\delta |^2)
\end{equation}
and
\begin{equation}
P(1_-(t))=\frac{1}{2} (|\alpha |^2 + |\gamma |^2)
\end{equation}
Except in the cases $|\beta|^2 + |\delta |^2 =1$ and $|\alpha |^2
+ |\gamma |^2= 1$ the information encoded by an observer on side $2$ (using adequate
interactions) as $|\beta|^2 + |\delta
|^2 $ and $|\alpha |^2 + |\gamma |^2 $ can be transmitted to side
$1$. If, moreover, we take $\alpha =1+ \alpha _o \theta (T), \beta
=\beta _o \theta (T),..$ with $\theta (T)$ the Heaviside function
($\theta (T)=0$ for $t<T$ and $\theta (T)=1$ for $t \geq T$) and
$T$ is chosen large enough that information can be transmitted in
a superluminal way.

This result can sound strange because we are accustomed to demand
quantum evolutions to be unitary in order to preserve
probabilities. However, as we shall show in the next subsection,
the non-unitarity is an "effective" characteristic of the
interferometric experiments when we disregard the particles
absorbed by the screen.

\subsection{Effective non-unitarity}

We show in this subsection that the dynamics of a particle
passing through a slit is non-unitary in an effective way because
of the particles absorbed by the screen surrounding the slit. The
situation is similar to that found in nuclear physics in
scattering problems when some of the incident particles are
captured by the target. In a phenomenological way the problem is
described by a complex potential, for instance the optical
potential \cite{nuc}. The non-unitarity associated with this
complex potential leads to the non conservation of probability
representing the capture of particles.

Let us consider now the problem of the diffraction of a particle
by a slit. The mathematical description of interferometry and
diffraction is particularly simple in the path
integration formalism \cite{Fey}. As the probability is conserved
if and only if the evolution is unitary, to demonstrate the non-
unitarity it suffices to show that the probability is not
conserved. The condition for the conservation of probability in
the path integral approach is
\begin{equation}
\int K^* ({\bf x_f}, t_f;{\bf x_i}', t_i ) K ({\bf x_f}, t_f;{\bf
x_i}, t_i ) d^3 {\bf x_f} = \delta ^3 ({\bf x_i}' - {\bf x_i} )
\end{equation}
where $K ({\bf x_f}, t_f;{\bf x_i}, t_i )$ is the kernel for a
particle going from $({\bf x_f}, t_f)$ to $({\bf x_i}, t_i ) $
(Eq. (4-37) of Ref. \cite{Fey}).

The arrangement we consider in this subsection consists of a
source which emits particles that impinge on a screen with a slit.
Some particles are stopped or absorbed by the screen and the rest
go through the slit being scattered and resulting in the usual
diffraction pattern after the screen. The problem is essentially
bidimensional with $x$ and $y$ representing the coordinates
parallel and perpendicular to the screen in the plane of source
and screen. The source is at the point $x=y=0$ and the width of
the slit is $2b$. The condition for the conservation of
probability becomes
\begin{eqnarray}
\int _{-\infty }^{\infty } d {\bf x_f} \int _{-\infty }^{\infty
} d {\bf y_f}  K^* ({\bf x_f}, {\bf y_f}, t_f; {\bf x_i}', {\bf
y_i}' t_i ) \times \nonumber \\
K ({\bf x_f}, {\bf y_f}, t_f;{\bf x_i}, {\bf y_i}, t_i )  =
\delta ({\bf x_i}' - {\bf x_i} ) \delta ({\bf y_i}' - {\bf y_i}
)
\label{eq:ncin}
\end{eqnarray}
In the path integral formalism the problem of diffraction of a
particle through a slit, which constrains its motion, is
approached by breaking the path into two successive free particle
motions, scattering the slit the particle from one to the other
free particle evolution \cite{Fey}. The kernel of a free particle
is:
\begin{equation}
K(a',a) = \frac{m}{2 \pi i \hbar (t_{a'} -t_a)} exp \left(
\frac{im((x_{a'}-x_a)^2 +(y_{a'}-y_a)^2 )}{2\hbar (t_{a'} -t_a)
} \right)
\label{eq:nsei}
\end{equation}
It is simple to check by direct substitution that (\ref{eq:nsei})
fulfills Eq. (\ref{eq:ncin}).

When the slit is present we have $K=K_x K_y$ with
\begin{equation}
K_x ({\bf x_f}, t_f;{\bf x_i}, t_i )  = \int _{-b}^b dx_c
exp \left( \frac{im(x_c-x_i)^2}{2\hbar (t_c -t_i) } \right)
exp \left( \frac{im(x_f-x_c)^2}{2\hbar (t_f -t_c) } \right)
\label{eq:nsie}
\end{equation}
where $x_c$ is the coordinate of the slit.

On the other hand $K_y$ is the kernel of a free particle since in the $y$-
direction the particle is not constrained. Consequently, the
integral on $y$ of $K_y$ in (\ref{eq:ncin}) gives $\delta (y_i
' - y_i)$.

Note that we have removed the factors before the exponential
because they are unessential for the final result. Introducing
Eq. (\ref{eq:nsie}) into Eq. (\ref{eq:ncin}) we have
\begin{eqnarray}
\int _{- \infty}^{\infty } dx_f \int _{-b}^{b} dx_c \int _{-
b}^{b} dx' _c  exp \left( \frac{im}{2\hbar } \left( \frac{x_c^2 -
(x '_c)^2 -2x_f(x_c - x' _c)}{t_f -t_c } \right) \right)
\nonumber \\
\times exp \left( \frac{im}{2\hbar } \left( \frac{x_c^2 +x_i^2 -
(x '_c)^2 -2x_i x_c +2 x' _c x' _i}{t_c -t_i } \right) \right)
\label{eq:noch}
\end{eqnarray}
First, we evaluate the integral on $x_f$
\begin{equation}
\int _{- \infty}^{\infty } dx_f exp \left( \frac{-im(x_c-x' _c)
x_f}{t_c -t_i } \right) = \delta \left( \frac{m(x' _c-x _c) }{t_f
-t_c } \right)
\end{equation}
Using this result we can evaluate Eq. (\ref{eq:noch}):
\begin{eqnarray}
\int _{-b}^{b} dx_c exp \left( \frac{im}{2\hbar } \left(
\frac{x_i^2 -(x '_i)^2 -2x_c(x_i - x' _i)}{t_c -t_i } \right)
\right) =  \\
exp \left( \frac{im}{2\hbar } \left( \frac{x_i^2 - (x' _i)^2 }
{t_c -t_i } \right) \right) \int _{-b}^b dx_c exp \left( \frac{-
im}{\hbar } \left( \frac{(x_i - x' _i)x_c } {t_c -t_i } \right)
\right) \neq \delta (x_i - x' _i )
\nonumber
\end{eqnarray}
We can only obtain the result $\delta (x_i - x' _i ) $ in the
limit $|b| \rightarrow \infty $.

We conclude that the probability is not conserved and the
evolution is not unitary. Clearly, the loss of probability is
associated with the absorption of particles (paths) by the screen.
These particles cannot be detected after the screen, and the
number of particles arriving to the detectors is smaller than the
number of emitted particles at the source. Thus the dynamics is
non-unitary in this restricted sense. If we would include in the
calculations the stopped particles we would recover the unitarity.

\section{Mechanisms preserving the coexistence}

We show in this Section that in addition to the dynamic condition
discussed in Sect. 2 other types of conditions must be taken into account
in order to decide if signaling is possible.

Let us imagine that we try to use the arrangement of Sect. 2 to transmit superluminal signals. In order to
obtain a detection pattern sufficiently clear from which to read
the information$"\phi "$ we need a minimum number of detected
particles $N$, which must be determined in every possible
experiment (depending on the type of detector, incident
particle...). We denote by $L$ the separation between the slit $C$
(where the element producing the phase change $\phi $ is placed)
and the detector of particle 1. Finally, $T$ is the time delay
between the arrival to the detector of the first and the last of
the $N$ particles. Obviously, the condition for the existence of
superluminal communication is $T<L/c$ with $c$ the light speed.
This condition can be rewritten in function of the mean time
between the emission of two entangled pairs by the source, $\tau
$, becoming the above expression:
\begin{equation}
\tau < \frac{L}{Nc}
\label{eq:ast}
\end{equation}
When the source can emit pairs of particles fulfilling this
condition we can transmit the information $" \phi "$ in a
superluminal way.

We can estimate the order of magnitude of $\tau $ by comparison
with standard one-particle interferometry. Typical values are
$N=7.000$ and $L=50 cm$, that would correspond to $L/Nc \approx
10^{-13} seconds$. This value is extremely short when compared to the
typical emission ratio of the usual sources of single particles,
around $10^{-3} seconds$.

In conclusion, the dynamic signaling plus the condition
(\ref{eq:ast}) leads to the possibility of signaling. On the other hand,
when we have $\tau \geq L/Nc$ the no-signaling condition is recovered.

The above considerations provide an example of a condition
preventing the possibility of signaling. They are an example of a
physical mechanism preserving the coexistence of Quantum Mechanics
and Special Relativity. It is not the only possible mechanism. For
instance, in the above analysis we have used the crude approximation
of an instantaneous response of the detector after the arrival of
the particles. It is by now well-known that this response time of
the detector is finite. In particular, in Ref. \cite{Ste} it has
been experimentally demonstrated that in optical mediums where the
group velocity exceeds the speed of light the detection of the
information encoded in the pulse takes a longer time, being
recovered in this way the relativistic limit for information
transmission. Probably, a complete non-dynamic condition for
no-signaling must include simultaneously considerations about the
emission and detection processes. See also Ref. \cite{Rub} for
related considerations in the context of superluminal behavior
associated with non-local propagators.

We conclude that there are physical mechanisms of a non-dynamic
type that could avoid the signaling, although from the dynamic
point of view it is in principle possible.

\section{Discussion}

We have analyzed in this paper a new type of non-locality different
from the usual one based on Bell-type correlations. The system
considered is a gedanken two-particle two-slit interference
arrangement with completely entangled states. Based on previous
results \cite{SPRa, Spa, Spl}, it is simple to show that, in
principle, a phase change in one of the particles can modify the
distant one-particle detection pattern of the companion particle.
This non-local effect could, in principle, be observed (in the
absence of other inhibiting mechanisms) detecting only the particles
in one side of the arrangement without having any information of the
results of the detection process on the other side, or even without
detecting at all that particles. This differentiates the arrangement
from the usual one in Bell-type experiments, where the nonlocal
effects are determined comparing the results of detections at both
sides of the arrangement (avoiding in this well-known way the
possibility of signaling).

The effects associated with this non-locality are extremely small
and, probably, unobservable. However, their importance can be
large from the foundational point of view. For instance, they show
interesting resemblances with the analysis of Hegerfeldt of the
acausal behavior of particles initially localized.

We have also introduced the distinction between signaling and
dynamic signaling. The second one refers to the possibility of
superluminal information transmission when only the dynamic
processes (evolution of the wave function) are taken into account.
It is a necessary condition. However, the possibility of
experimentally detecting signaling in a given arrangement also
requires from additional conditions related to the emission and
detection processes. Then signaling refers to the complete
evolution of the system, including the emission, subsequent
dynamical evolution and, finally, the detection.

In our example, where the dynamic signaling is in principle
possible, the no-signaling condition can be preserved by the
constrains on the ratio of emission of entangled pairs. As shown
experimentally in Ref. \cite{Ste} the general no-signaling
condition must also take into account the detection process.

The above results show that the dynamic no-signaling (the only
usually considered) is not a fundamental characteristic of
quantum dynamics and cannot be included between the basic
postulates of Quantum Theory. On the other hand, the possibility
of including the more general condition of no-signaling remains
an open question. However, it becomes clear from our analysis
that this possibility would require from a more general framework,
including emission and detection processes, than the one usually considered.

Another interesting question, closely related with the above one,
is the possibility of deducing the constrains on the emission
ratio or the detection properties from a purely quantum basis.
This would require from the quantum study of the behavior of very
general types of sources of entangled particles and detectors.

Finally, we remark that the constrain on the ratio of emission of
entangled particles strongly remembers the thermodynamic
formalism, where some general principles stating
the impossibility of some processes (perpetual motion of first
and second kind) play a fundamental role in the foundations of the
theory.

In conclusion, we have shown the existence of dynamic signaling
scenarios. However, we have remarked the existence of
non-dynamic mechanisms that could prevent the superluminal
transmission of information in these scenarios preserving the
coexistence of Quantum Mechanics and Special Relativity.

\end{document}